\documentstyle[aas2pp4,epsf]{article}
\lefthead{KASPI ET AL.}
\righthead{NGC 4151 MONITORING II. OPTICAL OBSERVATIONS}

\def\OIII{O\,{\sc iii}}
\def\Civ{C\,{\sc iv}}
\def\Heii{He\,{\sc ii}}

\begin{document}

\vspace{5cm}

\title{  
MULTIWAVELENGTH OBSERVATIONS OF \\
SHORT TIME-SCALE VARIABILITY IN NGC~4151. \\
II. OPTICAL OBSERVATIONS\altaffilmark{1}
\altaffiltext{1}{This paper is dedicated to the memory of K.K.~Chuvaev
who passed away in the course of this work.}
}



\author{
S.~Kaspi,\altaffilmark{2}
\altaffiltext{2}{School of Physics and Astronomy and the Wise
Observatory, The Beverly and Raymond Sackler Faculty of Exact Sciences,
Tel-Aviv University, Tel-Aviv 69978, Israel.}
D.~Maoz,\altaffilmark{2}
H.~Netzer,\altaffilmark{2}
B.~M.~Peterson,\altaffilmark{3} 
\altaffiltext{3}{Department of Astronomy, The Ohio State University,
174 West 18th Avenue, Columbus, OH 43210.} \\
T.~Alexander,\altaffilmark{2}
A.~J.~Barth,\altaffilmark{4}
\altaffiltext{4}{Department of Astronomy, University of California at
Berkeley, Berkeley, CA  94720.}
R.~Bertram,\altaffilmark{3,5}
\altaffiltext{5}{Mailing address: Lowell Observatory, 1400 West Mars
Hill Road, Flagstaff, AZ 86001.}
F.~-Z.~Cheng,\altaffilmark{6}  \\
\altaffiltext{6}{Center for Astrophysics, University of Science and
Technology, Hefei, Anhui, People's Republic of China.}
K.~K.~Chuvaev,\altaffilmark{7,8}
\altaffiltext{7}{Crimean Astrophysical Observatory, P/O Nauchny, 334413
Crimea, Ukraine.}
\altaffiltext{8}{Deceased, 1994 November 15.}
R.~A.~Edelson,\altaffilmark{9}
\altaffiltext{9}{Department of Physics and Astronomy, University of Iowa,
Iowa City, IA 52242.}
A.~V.~Filippenko,\altaffilmark{4} \\
S.~Hemar,\altaffilmark{2}
L.~C.~Ho,\altaffilmark{4}
O.~Kovo,\altaffilmark{2}
T.~Matheson,\altaffilmark{4}
R.~W.~Pogge,\altaffilmark{3} \\
B.~-C.~Qian,\altaffilmark{10}
\altaffiltext{10}{Shanghai Observatory, Chinese Academy of Sciences,
People's Republic of China.}
S.~M.~Smith,\altaffilmark{3}
R.~M.~Wagner,\altaffilmark{3,5} \\
H.~Wu,\altaffilmark{11}
\altaffiltext{11}{Beijing Astronomical Observatory, Chinese Academy of
Sciences, Beijing 100080, People's Republic of China.}
S.~-J.~Xue,\altaffilmark{6}
and Z.~-L.~Zou \altaffilmark{11}
}

\vspace{0.5cm}

\centerline{To appear in the $ApJ$, October 20, 1996 issue, Vol. 470}
\vspace{0.1cm}

\centerline{Preprint Series No. 96/72}

\vspace{-1.2cm}

\begin{abstract}
We present the results of an intensive ground-based spectrophotometric
monitoring campaign of the Seyfert galaxy NGC~4151 for a period of over
two months, with a typical temporal resolution of one day. Light curves
for four optical continuum bands and the H$\alpha$ and H$\beta$
emission lines are given. During the monitoring period, the continuum
at 6925~\AA\ varied by $\sim$17\% while the continuum at
4600~\AA\ varied by $\sim$35\%, with larger variations in the near UV.
The wavelength dependence of the variation amplitude also extends into
the far UV. The dependence in the 2700$-$7200~\AA\ range can be
explained by the different relative starlight contributions at
different wavelengths, but the large variability at 1275~\AA\ cannot be
explained in this way. The continuum variability timescale is of order
13 days and is similar at all optical wavelength bands. No evidence for
a time lag between the optical continuum and the UV continuum and
emission lines was found.  The H$\alpha$ emission line flux varied by
about 12\% with a gradual rise throughout the campaign.  Its cross
correlation with the continuum light curve gives a lag of $0-2$ days.
The variations in the H$\beta$ emission line flux are about 30\% and
lag the continuum by 0$-$3 days. This is in contrast to past results
where a time lag of 9$\pm$2 days was found for both emission lines.
This may be due to a different variability timescale of the
{\em{ionizing}} continuum, or to a real change in the BLR gas
distribution in the 5.5 years interval between the two campaigns.
\end{abstract}


\section{Introduction}

The Seyfert~1 galaxy NGC~4151 is one of the best-studied active
galactic nuclei (AGN) due to its brightness and variability
properties.  It has been studied at many wavelengths and its
characteristics are well known (e.g., Peterson 1988). Several
monitoring campaigns have shown variability timescales from a few hours
in the hard X-rays (Yaqoob et al.  1993), to a few days in the
ultraviolet (e.g.,  Clavel et al.  1990) and the optical (e.g., Maoz et
al. 1991), and several months in the IR (Prestwich, Wright, \& Joseph
1992).

NGC~4151 was selected by the AGN Watch consortium as a prime target for
an intensive spectroscopic multiwavelength monitoring campaign. The
campaign took place for two weeks in 1993 December using the
$CGRO,\ ASCA,\  ROSAT,$ and $IUE$ satellites, and many ground-based
telescopes. The Crenshaw et al. paper in this issue (hereafter Paper I)
describes the UV results from $IUE$. This paper describes the
observations and results of the ground-based optical campaign. The
high-energy results ($GRO,\ ASCA,\  ROSAT$) are described by Warwick et
al. in this issue (Paper III), and a multiwavelength comparison is
given by Edelson et al. (Paper IV).

A primary goal of AGN monitoring has been to determine the size of the
broad-line region (BLR; see Peterson 1993 for a review). Of the many
ground-based variability studies of NGC~4151, the most intensive have
been those of Antonucci \& Cohen (1983) and Maoz et al. (1991).
Antonucci \& Cohen monitored NGC~4151 at approximately monthly
intervals for over a year.  They found that the continuum and broad
lines varied on timescales shorter than their temporal resolution, and
deduced a BLR radius of less than $\sim$30 lt-days. 
For these data Gaskell \& Sparke (1986) reported an H$\beta$ lag
of $0-7$ days and H$\gamma$ lag of $5-9$ days.
These observations were also analyzed by Peterson \& Cota (1988) who
combined them with their own data and arrived at a BLR size of 6$\pm$4
lt-days. The Maoz et al. (1991) monitoring campaign lasted for a period
of 8 months with a mean sampling interval of 4 days.  Cross
correlation, deconvolution, and modeling applied to the data indicated
a BLR size of 9$\pm$2 lt-days. The $IUE$ monitoring campaign described
by Clavel et al.  (1990) yielded a characteristic timescale of $4\pm 3$
days from the response of the C\,{\sc iv} $\lambda1549$ and Mg\,{\sc
ii} $\lambda2798$ emission lines to the UV continuum.

In this paper we present results of two months of monitoring of
NGC~4151 with a time resolution of about one day. In \S 2 we describe
the ground-based observations together with their reduction and
calibration, and present light curves for the continuum and the
H$\alpha$ and H$\beta$ emission lines. In \S 3 we carry out a
time-series analysis of the data and briefly discuss the results. A
summary is given in \S 4.

\section{Observations}

\subsection{Data and Reduction}

The main effort of the ground-based monitoring campaign was carried out
at the Wise Observatory in Israel, using the 1m telescope. The last two
hours of each night, for a period of over two months starting on 1993
November 14, were dedicated to the monitoring of NGC~4151.
Spectroscopic observations were performed with the Faint Object
Spectrographic Camera (Kaspi et al. 1995), using a
$10^{\prime\prime}$-wide slit and a 600 line/mm grism, giving a
dispersion of 2~\AA\ per pixel. The spectral resolution was determined
by the seeing, which was 2 to 3 arcsec, and combined with the
instrument spatial scale of 0.9 arcsec/pixel the spectral resolution
was $\sim$5~\AA . Two different slits were used every night, each
located at a different position in the telescope's focal plane. One
slit produced spectra in the range $\sim$~4150$-$6050~\AA\ (hereafter
the ``B-side'') and the other covered the range
$\sim$~5060$-$6990~\AA\ (the ``R-side''). With this setup we were able
to monitor both the H$\alpha$ and the H$\beta$ emission lines, while a
large part of the AGN continuum between them was observed through both
slits.  This produced two independent measurements of the
5100$-$6000~\AA\ continuum, since the object was centered in each slit
separately.

The spectrograph was rotated to include, together with the nucleus of
NGC~4151, a field star (``star 1'' of Penston, Penston, \& Sandage
1971, P.A. $=156.3^{\circ}$) that served as a local standard. This
technique, of using a local comparison star, is described in detail by
Maoz et al. (1990) and Maoz et al. (1994), and produces high relative
spectrophotometric accuracy.

For each of the B-side and R-side setups, we tried to obtain two or
more consecutive exposures per night, with typical integration times of
15 min per exposure. The data were reduced using standard IRAF routines
with an extraction window of $\sim~13^{\prime\prime}$.  The
NGC~4151/star ratios of consecutive exposures were compared to test for
systematic errors. The ratios almost always reproduced to 0.2\%$-$1.2\%
at all wavelengths. The few nights where the ratios between exposures
differed by more than $3.5\%$ were discarded.  (Only three nights were
with ratios between 1.2\% and 3.5\%.) The NGC~4151 and the star spectra
from the consecutive exposures were co-added to improve the
signal-to-noise ratio, and the resultant NGC~4151 spectrum was divided
by an absorption-line-free, interpolated and smoothed version of the
co-added stellar spectrum. This removes atmospheric absorption from the
Seyfert spectrum and provides the relative flux calibration.  This
procedure resulted in one spectrum per setup per night. Absolute
photometric calibration was achieved by multiplying the NGC~4151/star
ratio by a smoothed flux-calibrated spectrum of the comparison star,
obtained on photometric nights using known spectrophotometric
standards.

\begin{deluxetable}{ccc}
\tablecolumns{3}
\tablecaption {Flux scale factors \label {phiandG}}
\tablewidth{0pt}
\tablehead{
\colhead{} & \colhead{Point-Source} & \colhead{Extended Source} \nl
\colhead{Data Set} & \colhead{Scale Factor $\varphi$} & \colhead{Correction $G$\tablenotemark{a}} }
\startdata
OSU      &  0.9558 $\pm$ 0.0027  &  -0.758 $\pm$ 0.018  \nl
Lick     &  0.9349 $\pm$ 0.0043  &  -1.103 $\pm$ 0.023  \nl
Wise     &          1            &             0        \nl
\enddata
\tablenotetext{a}{In units of $10^{-14}$ ergs s$^{-1}$ cm$^{-2}$ \AA $^{-1}$}
\end{deluxetable}

The second data set was obtained by the Ohio State University (OSU)
group with the Perkins 1.8m telescope at the Lowell Observatory.
NGC~4151 was observed for 10 nights between 1993 December 2 and 11, UT,
with the Boller \& Chivens spectrograph through a 5 arcsec slit at
P.A.=90$^{\circ}$.  Many exposures were taken each night with
integration times of $2-3$ min each, with a total integration time of
several hours (in order to study the variations on timescales of
minutes). The spectra covered the wavelength range
$\sim$~4480$-$5660~\AA . The data were reduced using standard IRAF
routines.  The individual spectra were co-added to produce three
spectra for each night, which, after checking for agreement of 1\% or
less, were added into one spectrum per night.

The third data set that will be presented here was obtained with the
Kast double spectrograph (Miller \& Stone 1993) on the Shane 3 m
reflector at Lick Observatory, using a 4 arcsec slit which was aligned
along the parallactic angle. This set consists of 3 epochs. Seven other
data sets were obtained by other AGN watch members covering anywhere
between 1 and 8 nights. These will be discussed in the following
section.

\subsection{Calibration}

A project of this scope requires the intercalibration of the various
data sets into a single consistent set.  The method used for this
intercalibration is based on the [\OIII ] $\lambda\lambda$4959, 5007
narrow emission lines, by requiring all spectra to have the same flux
in these lines. There are various ways of achieving this objective. The
one adopted here is based on Peterson et al. (1994).

The Wise data set, being the largest one, was chosen as the reference
data set. The [\OIII ] $\lambda\lambda$4959, 5007 emission lines fluxes
were measured between the observed wavelengths of 4952$-$5051~\AA , by
summing the measured flux above a straight line passing between the
average flux in the 6~\AA\ interval on each side of this wavelength
range. This kind of measurement was determined by Baribaud \& Alloin
(1990) to be optimal for the [\OIII ] emission lines.

A second scaling method used is described by van Groningen \& Wanders
(1992). The method finds the optimum scaling factor, wavelength shift,
and convolution factor of one spectrum with respect to a reference
spectrum, by slowly varying these parameters until the residuals of one
or more non-variable narrow lines in the difference between the two
spectra are minimized. This resulted in spectra and light curves which
confirmed but did not improve (in terms of the internal scatter within
one data set) the results of the former method.

The mean flux of the two [\OIII ] emission lines through the Wise
apertures was found to be ${\cal F}_{[O~III]}=1.575\times 10^{-11}$
ergs s$^{-1}$ cm$^{-2}$, with a scatter of $1.1\%$. (This is consistent
with the Antonucci \& Cohen (1983) result of
$(1.19\pm~0.06)~\times~10^{-11}$ ergs s$^{-1}$ cm$^{-2}$ for the [\OIII
] $\lambda 5007$ flux.) We consider this to be the accuracy of the
comparison-star method. As a test, all B-side spectra were scaled so
that their measured ${\cal F}_{[O~III]}$ agree with the above flux.  No
obvious differences appear in the Wise light curves before and after
the [\OIII ] scaling, demonstrating the reliability of the
comparison-star method.  An attempt to scale all R-side spectra to the
same [S\,{\sc ii}] $\lambda\lambda$6716, 6731 emission line fluxes gave
similar results.  A comparison of the overlapping continuum bands
between the R-side and the B-side spectra shows an agreement of about
1\% for almost all epochs.

All other data sets were scaled to match their [\OIII ] line fluxes to
the Wise ${\cal F}_{[O\,III]}$ by multiplying each spectrum by a factor
$\frac{{\cal F}_{[O\,III]}}{F([O\,III])_{obs}}$, where
$F([O\,III])_{obs}$ is the measured [\OIII ] line flux of the
spectrum.  Light curves were then measured for the H$\beta$ line and
the 5100$-$5150~\AA\ continuum.  These two light curves, which were
obtained for each data set, were then intercalibrated with the Wise
data set.

The intercalibration method (Peterson et al. 1994) is as follows. First,
a point-source correction factor $\varphi$ was defined by the equation
\begin{equation}
F(H\beta) = \varphi F(H\beta )_{obs} \ ,       \label{phi-def}
\end{equation}
where  $F(H\beta )_{obs}$ is the H$\beta$ line flux measured for each
spectrum after scaling its [\OIII ] line flux to agree with the Wise
${\cal F}_{[O~III]}$, and $F(H\beta )$ is the Wise H$\beta$ flux from an
observation that is close in time (see below). The $\varphi$ factor
accounts for the fact that different apertures resulted in different
amounts of light loss for the given point-spread function (which
describes the surface-brightness distribution of both the broad lines
and the AGN continuum source) and the partially extended narrow-line
region.

An additive correction was applied to allow for the different amounts
of starlight admitted by different apertures. This correction, $G$, was
defined by the equation
\begin{equation}
F_{\lambda}(5100) = \varphi F_{\lambda}(5100)_{obs} -  G  \ ,
                                                               \label{G-def}
\end{equation}
where $F_{\lambda}(5100)_{obs}$ is the continuum measured in the
observed wavelength range 5100$-$5150~\AA\ after the spectrum was
scaled to have the Wise [\OIII ] line flux, and $F_{\lambda}(5100)$ is
the contemporaneous Wise result.

A problem unique to this monitoring project is that NGC~4151 could be
observed by ground-based telescopes for only about two hours at the end
of each night. (Due to $ROSAT$ constraints, the campaign on this
object, which has a right ascension of $\sim 12$ hours, was executed in
December.) Thus, no pairs of nearly simultaneous observations from
different observatories (to determine the values of $\varphi$ and $G$)
could be found. This limits our ability to intercalibrate the various
data sets and is different from previous monitoring campaigns (e.g.,
NGC~5548) where objects could be observed all night, and the time
resolution was a few days.

Given the above difficulty, we have taken the alternative approach of
interpolation.  The reference Wise light curves of H$\beta$ and
$F_{\lambda}(5100)$ were interpolated linearly, and for every spectrum
in the other data sets the values of $F(H\beta )$ and
$F_{\lambda}(5100)$ at their precise date and time were found, and used
for defining $\varphi$ and $G$ (from eqs.~\ref{phi-def} and
\ref{G-def}). The $\varphi$ and $G$ of each data set were then
determined by averaging the values that were obtained for each spectrum
of a data set.  Finally, each spectrum of the given data set was
multiplied by the average $\varphi$ and subtracted by $G$, producing
the final calibrated set.

The intercalibration constants, $\varphi$ and $G$, for each data set
are given in Table~\ref{phiandG}. The $\varphi$ for the OSU data set is
in good agreement with its value given by Peterson et al. (1995), who
calculated it from the [\OIII ] $\lambda$5007 surface-brightness
\ distribution of \ NGC4151 \ through different apertures.

The intercalibration method can only be used for data sets with more
than two epochs.  Based on this criterion, several data sets had to be
discarded.  Also, all data sets with an internal accuracy worse than
3\% (i.e., with spectra from the same night differing from each other,
or from the Wise set, by more than 3\%) were discarded.  Given the
small variation of the source during the period of intensive monitoring
(1993 December 2$-$11), these data sets do not add useful information
and increase the overall noise.

In view of all these limitations, only the two large data sets from
Wise and OSU, as well as the smaller Lick data set, will be discussed
in the following sections.

\subsection{Line and Continuum Light Curves}

Fig.~\ref{lc}a and Table~\ref{con_lc} give mean continuum light
curves for four spectral bands: (a) 4560$-$ 4640~\AA\ (hereafter
4600~\AA ), (b) 5100$-$5150~\AA\ (average of the Wise B-side and
R-side, and several points from the OSU and Lick sets, hereafter
5125~\AA ), (c) 6170$-$6230~\AA\ (hereafter 6200~\AA ), and (d)
6900$-$6950~\AA\ (hereafter 6925~\AA ). All wavelengths are in the
observed reference frame.  All four continuum light curves show the
same feature of small ``bumps'' superposed on a gradual rise of the
flux throughout the whole monitoring period. Examination of these light
curves shows that the optical continuum changed (from a minimum at
JD=2449315.6 to a maximum at JD=2449357.6) by about $35\%$ at 4600~\AA
\ and $17\%$ at 6925~\AA\ ($\frac{F_{max}}{F_{min}}-1$). This can also
be seen in Fig.~\ref{spectrum} (bottom panel): the r.m.s. of the Wise
light curves is $\sim 10\%$ at 4600~\AA\ whereas at 6925~\AA\ it is
$\sim 5\%$.

The H$\beta$ flux was measured from each frame of the Wise B-side, OSU,
and Lick sets, between the observed wavelengths 4780$-$4950~\AA , by
summing the measured flux above a straight line passing between the
average flux in the intervals 4560$-$4640~\AA\ and 5100$-$5150~\AA .
The H$\alpha$ emission line was measured from the Wise R-side and Lick
spectra between 6415$-$ 6710~\AA , and its underlying continuum was
measured between the average flux of two wavelength ranges
6170$-$6230~\AA\ and 6890$-$6970~\AA . Fig.~\ref{lc}b and
Table~\ref{con_lc} show the H$\beta$ light curve, and Fig.~\ref{lc}c
and Table~\ref{con_lc} show the H$\alpha$ light curve. Both H$\alpha$
and H$\beta$ show a gradual rise in flux through the monitoring period,
which amounts to $\sim$13\% in H$\alpha$ and to $\sim 30\%$ in
H$\beta$. While in the H$\beta$ light curve there are several features
which resemble the continuum light curve (Fig.~\ref{lc}a), such
features are barely visible in the H$\alpha$ light curve.

\begin{deluxetable}{ccccccc}
\tablecolumns{7}
\scriptsize
\tablecaption {Light curves \label {con_lc}}
\tablewidth{0pt}
\tablehead{
\colhead{JD} & \colhead{4600~\AA} & \colhead{5125~\AA} & \colhead{6200~\AA} & \colhead{6925~\AA}  & \colhead{H$\alpha$} & \colhead{H$\beta$}
}
\startdata
305.6\phm{$^{*}$} & \phn 8.44$\pm$0.09 &  7.22$\pm$0.07 &  6.07$\pm$0.06 &  5.86$\pm$0.06 & 27.36$\pm$0.28 &  6.39$\pm$0.09 \nl
306.6\phm{$^{*}$} & \phn 8.72$\pm$0.05 &  7.51$\pm$0.04 &  6.29$\pm$0.03 &  6.16$\pm$0.03 & 27.20$\pm$0.14 &  6.44$\pm$0.06 \nl
307.6\phm{$^{*}$} & \phn 8.88$\pm$0.05 &  7.44$\pm$0.04 &  \nodata       &  \nodata       &  \nodata       &  6.73$\pm$0.07 \nl
308.6\phm{$^{*}$} & \phn 8.91$\pm$0.03 &  7.48$\pm$0.03 &  6.29$\pm$0.03 &  6.11$\pm$0.03 & 27.47$\pm$0.12 &  6.74$\pm$0.06 \nl
309.1$^{\dagger}$ & \phn 8.75$\pm$0.09 &  7.51$\pm$0.08 &  6.21$\pm$0.06 &  6.08$\pm$0.06 & 27.43$\pm$0.22 &  6.70$\pm$0.03 \nl
309.6\phm{$^{*}$} & \phn 8.76$\pm$0.07 &  7.48$\pm$0.06 &  6.30$\pm$0.02 &  6.14$\pm$0.02 & 27.62$\pm$0.09 &  6.72$\pm$0.09 \nl
310.6\phm{$^{*}$} & \phn 8.68$\pm$0.08 &  7.39$\pm$0.07 &  6.23$\pm$0.06 &  6.06$\pm$0.06 & 27.94$\pm$0.25 &  7.09$\pm$0.15 \nl
311.6\phm{$^{*}$} & \phn 8.55$\pm$0.05 &  7.41$\pm$0.05 &  6.28$\pm$0.03 &  6.10$\pm$0.03 & 27.81$\pm$0.15 &  6.50$\pm$0.06 \nl
312.6\phm{$^{*}$} & \phn 8.44$\pm$0.16 &  7.29$\pm$0.09 &  \nodata       &  \nodata       &  \nodata       &  7.14$\pm$0.35 \nl
313.6\phm{$^{*}$} & \phn 8.35$\pm$0.07 &  7.23$\pm$0.06 &  6.18$\pm$0.04 &  6.02$\pm$0.04 & 27.55$\pm$0.20 &  6.60$\pm$0.10 \nl
314.6\phm{$^{*}$} & \phn 8.37$\pm$0.02 &  7.30$\pm$0.02 &  6.20$\pm$0.01 &  6.02$\pm$0.01 & 27.77$\pm$0.05 &  6.81$\pm$0.05 \nl
315.6\phm{$^{*}$} & \phn 8.04$\pm$0.23 &  7.33$\pm$0.19 &  6.23$\pm$0.04 &  6.02$\pm$0.04 & 27.53$\pm$0.17 &  6.87$\pm$0.32 \nl
316.6\phm{$^{*}$} & \phn 8.20$\pm$0.09 &  7.33$\pm$0.07 &  6.26$\pm$0.02 &  5.96$\pm$0.02 & 27.91$\pm$0.10 &  6.58$\pm$0.11 \nl
317.6\phm{$^{*}$} & \nodata            &  7.41$\pm$0.08 &  6.25$\pm$0.06 &  6.02$\pm$0.06 & 27.66$\pm$0.29 &  \nodata       \nl
318.6\phm{$^{*}$} & \phn 8.96$\pm$0.12 &  7.62$\pm$0.08 &  6.40$\pm$0.07 &  6.17$\pm$0.06 & 27.88$\pm$0.28 &  6.29$\pm$0.20 \nl
319.6\phm{$^{*}$} & \phn 9.01$\pm$0.05 &  7.81$\pm$0.16 &  6.38$\pm$0.13 &  6.20$\pm$0.12 & 27.85$\pm$0.56 &  6.60$\pm$0.08 \nl
320.6\phm{$^{*}$} & \phn 9.42$\pm$0.10 &  8.20$\pm$0.09 &  6.61$\pm$0.02 &  6.20$\pm$0.02 & 27.75$\pm$0.10 &  6.98$\pm$0.11 \nl
321.6\phm{$^{*}$} & \phn 9.58$\pm$0.18 &  \nodata       &  \nodata       &  \nodata       &  \nodata       &  \nodata       \nl
322.6\phm{$^{*}$} &  \nodata           &  8.16$\pm$0.03 &  6.84$\pm$0.03 &  6.46$\pm$0.03 & 28.14$\pm$0.13 &  \nodata       \nl
324.0$^{*}$       & \phn 9.96$\pm$0.10 &  8.19$\pm$0.08 &  \nodata       &  \nodata       &  \nodata       &  7.21$\pm$0.07 \nl
324.6\phm{$^{*}$} & 10.06$\pm$0.11     &  8.30$\pm$0.09 &  6.84$\pm$0.01 &  6.53$\pm$0.02 & 28.28$\pm$0.06 &  7.26$\pm$0.13 \nl
325.0$^{*}$       & 10.00$\pm$0.10     &  8.32$\pm$0.08 &  \nodata       &  \nodata       &  \nodata       &  7.20$\pm$0.07 \nl
325.6\phm{$^{*}$} & 10.07$\pm$0.06     &  8.41$\pm$0.04 &  6.86$\pm$0.02 &  6.58$\pm$0.03 & 28.17$\pm$0.10 &  7.24$\pm$0.07 \nl
326.0$^{*}$       & \phn 9.93$\pm$0.10 &  8.31$\pm$0.08 &  \nodata       &  \nodata       &  \nodata       &  7.08$\pm$0.07 \nl
326.6\phm{$^{*}$} & 10.20$\pm$0.07     &  8.56$\pm$0.06 &  \nodata       &  \nodata       &  \nodata       &  7.39$\pm$0.07 \nl
326.9$^{*}$       & 10.15$\pm$0.10     &  8.41$\pm$0.08 &  \nodata       &  \nodata       &  \nodata       &  7.38$\pm$0.07 \nl
327.6\phm{$^{*}$} & 10.15$\pm$0.07     &  8.47$\pm$0.05 &  6.62$\pm$0.08 &  6.38$\pm$0.08 & 28.43$\pm$0.34 &  7.22$\pm$0.08 \nl
327.9$^{*}$       & 10.12$\pm$0.10     &  8.44$\pm$0.08 &  \nodata       &  \nodata       &  \nodata       &  7.40$\pm$0.07 \nl
328.6\phm{$^{*}$} & \phn 9.96$\pm$0.08 &  8.30$\pm$0.09 &  6.85$\pm$0.07 &  6.63$\pm$0.07 & 28.91$\pm$0.30 &  7.37$\pm$0.10 \nl
328.9$^{*}$       & 10.07$\pm$0.10     &  8.36$\pm$0.08 &  \nodata       &  \nodata       &  \nodata       &  7.40$\pm$0.07 \nl
329.6\phm{$^{*}$} & \phn 9.91$\pm$0.13 &  8.37$\pm$0.07 &  6.75$\pm$0.02 &  6.71$\pm$0.04 & 28.47$\pm$0.08 &  7.51$\pm$0.32 \nl
329.9$^{*}$       & 10.07$\pm$0.10     &  8.36$\pm$0.08 &  \nodata       &  \nodata       &  \nodata       &  7.38$\pm$0.07 \nl
330.6\phm{$^{*}$} & \phn 9.87$\pm$0.11 &  8.23$\pm$0.09 &  6.62$\pm$0.06 &  6.47$\pm$0.06 & 28.09$\pm$0.26 &  7.27$\pm$0.11 \nl
331.0$^{*}$       & \phn 9.81$\pm$0.10 &  8.31$\pm$0.08 &  \nodata       &  \nodata       &  \nodata       &  7.33$\pm$0.07 \nl
331.6\phm{$^{*}$} & \phn 9.80$\pm$0.05 &  8.30$\pm$0.04 &  6.88$\pm$0.02 &  6.63$\pm$0.02 & 28.35$\pm$0.07 &  7.21$\pm$0.09 \nl
331.9$^{*}$       & \phn 9.77$\pm$0.10 &  8.33$\pm$0.08 &  \nodata       &  \nodata       &  \nodata       &  7.31$\pm$0.07 \nl
332.6\phm{$^{*}$} & \phn 9.69$\pm$0.13 &  8.15$\pm$0.11 &  6.77$\pm$0.03 &  6.51$\pm$0.03 & 28.45$\pm$0.12 &  7.32$\pm$0.13 \nl
333.0$^{*}$       & \phn 9.68$\pm$0.10 &  8.35$\pm$0.08 &  \nodata       &  \nodata       &  \nodata       &  7.30$\pm$0.07 \nl
333.6\phm{$^{*}$} & \phn 9.98$\pm$0.11 &  8.26$\pm$0.09 &  6.78$\pm$0.07 &  6.55$\pm$0.07 & 28.21$\pm$0.29 &  7.25$\pm$0.15 \nl
334.6\phm{$^{*}$} & \phn 9.75$\pm$0.04 &  8.26$\pm$0.03 &  6.88$\pm$0.02 &  6.62$\pm$0.02 & 29.25$\pm$0.10 &  7.31$\pm$0.07 \nl
335.5\phm{$^{*}$} & \nodata            &  8.33$\pm$0.09 &  6.75$\pm$0.07 &  6.51$\pm$0.07 & 28.99$\pm$0.30 &  \nodata       \nl
336.6\phm{$^{*}$} & 10.08$\pm$0.05     &  8.42$\pm$0.05 &  6.84$\pm$0.04 &  6.62$\pm$0.04 & 28.77$\pm$0.18 &  7.39$\pm$0.06 \nl
337.6\phm{$^{*}$} & 10.22$\pm$0.07     &  8.41$\pm$0.06 &  6.88$\pm$0.04 &  6.70$\pm$0.03 & 28.98$\pm$0.15 &  7.35$\pm$0.07 \nl
338.6\phm{$^{*}$} & 10.26$\pm$0.06     &  8.52$\pm$0.04 &  7.02$\pm$0.02 &  6.75$\pm$0.02 & 29.29$\pm$0.07 &  7.20$\pm$0.08 \nl
339.6\phm{$^{*}$} & 10.36$\pm$0.07     &  8.51$\pm$0.07 &  7.02$\pm$0.06 &  6.75$\pm$0.05 & 29.36$\pm$0.24 &  7.25$\pm$0.07 \nl
340.1$^{\dagger}$ & 10.36$\pm$0.10     &  8.50$\pm$0.08 &  7.01$\pm$0.07 &  6.74$\pm$0.07 & 28.88$\pm$0.24 &  7.34$\pm$0.03 \nl
340.6\phm{$^{*}$} & 10.24$\pm$0.10     &  8.42$\pm$0.08 &  6.98$\pm$0.04 &  6.65$\pm$0.04 & 28.70$\pm$0.16 &  7.55$\pm$0.11 \nl
341.6\phm{$^{*}$} & 10.11$\pm$0.12     &  8.31$\pm$0.09 &  6.99$\pm$0.07 &  6.69$\pm$0.07 & 29.13$\pm$0.30 &  7.36$\pm$0.19 \nl
342.6\phm{$^{*}$} & \phn 9.44$\pm$0.07 &  8.04$\pm$0.05 &  6.70$\pm$0.04 &  6.54$\pm$0.04 & 28.29$\pm$0.15 &  7.35$\pm$0.10 \nl
346.6\phm{$^{*}$} & \phn 9.29$\pm$0.30 &  7.70$\pm$0.25 &  \nodata       &  \nodata       &  \nodata       &  7.32$\pm$0.25 \nl
350.5\phm{$^{*}$} & \nodata            &  7.92$\pm$0.12 &  6.51$\pm$0.10 &  6.42$\pm$0.10 & 28.92$\pm$0.44 &  \nodata       \nl
352.6\phm{$^{*}$} & 10.30$\pm$0.17     &  8.33$\pm$0.13 &  6.89$\pm$0.07 &  6.60$\pm$0.07 & 28.91$\pm$0.30 &  7.38$\pm$0.19 \nl
353.6\phm{$^{*}$} & 10.48$\pm$0.08     &  8.50$\pm$0.06 &  \nodata       &  \nodata       &  \nodata       &  7.23$\pm$0.10 \nl
355.6\phm{$^{*}$} & 10.69$\pm$0.04     &  8.71$\pm$0.05 &  7.13$\pm$0.04 &  6.87$\pm$0.04 & 29.44$\pm$0.15 &  7.44$\pm$0.09 \nl
357.6\phm{$^{*}$} & 10.84$\pm$0.08     &  8.76$\pm$0.05 &  7.26$\pm$0.03 &  7.05$\pm$0.03 & 29.41$\pm$0.11 &  7.57$\pm$0.15 \nl
358.6\phm{$^{*}$} & 10.72$\pm$0.04     &  8.78$\pm$0.03 &  7.28$\pm$0.03 &  6.99$\pm$0.02 & 30.05$\pm$0.11 &  7.84$\pm$0.10 \nl
359.6\phm{$^{*}$} & 10.66$\pm$0.04     &  8.78$\pm$0.02 &  7.20$\pm$0.01 &  6.94$\pm$0.02 & 29.82$\pm$0.06 &  7.65$\pm$0.10 \nl
360.1$^{\dagger}$ & 10.66$\pm$0.11     &  8.70$\pm$0.09 &  7.28$\pm$0.07 &  6.95$\pm$0.07 & 30.09$\pm$0.24 &  7.76$\pm$0.08 \nl
362.6\phm{$^{*}$} & 10.36$\pm$0.11     &  8.62$\pm$0.09 &  7.19$\pm$0.02 &  6.91$\pm$0.02 & 29.83$\pm$0.08 &  7.91$\pm$0.14 \nl
366.5\phm{$^{*}$} & 10.62$\pm$0.31     &  8.54$\pm$0.25 &  7.14$\pm$0.10 &  6.90$\pm$0.10 & 29.83$\pm$0.42 &  8.05$\pm$0.26 \nl
369.6\phm{$^{*}$} & 10.03$\pm$0.12     &  8.34$\pm$0.10 &  6.96$\pm$0.04 &  6.74$\pm$0.04 & 29.48$\pm$0.19 &  7.85$\pm$0.17 \nl
372.6\phm{$^{*}$} & 10.78$\pm$0.07     &  8.60$\pm$0.05 &  7.07$\pm$0.03 &  6.84$\pm$0.03 & 29.87$\pm$0.13 &  7.99$\pm$0.11 \nl
373.6\phm{$^{*}$} & 10.67$\pm$0.13     &  8.69$\pm$0.10 &  \nodata       &  \nodata       &  \nodata       &  7.91$\pm$0.14 \nl
377.6\phm{$^{*}$} & \nodata            &  8.75$\pm$0.04 &  7.15$\pm$0.03 &  6.91$\pm$0.03 & 29.96$\pm$0.13 &  \nodata       \nl
384.6\phm{$^{*}$} & 11.14$\pm$0.04     &  8.89$\pm$0.05 &  7.21$\pm$0.04 &  6.97$\pm$0.04 & 30.10$\pm$0.16 &  8.17$\pm$0.10 \nl
403.4\phm{$^{*}$} & 10.57$\pm$0.06     &  8.57$\pm$0.04 &  7.12$\pm$0.04 &  6.99$\pm$0.04 & 30.71$\pm$0.16 &  8.43$\pm$0.12 \nl
404.4\phm{$^{*}$} & 10.18$\pm$0.09     &  8.50$\pm$0.09 &  7.00$\pm$0.07 &  6.92$\pm$0.07 & 31.34$\pm$0.32 &  8.38$\pm$0.14 \nl
\enddata
\tablenotetext{}{$^{*}$ OSU data.  \ \ $^{\dagger}$ Lick data.}
\tablenotetext{1}{Observation Julian Date $-$ 2449000, rounded to a tenth of a day.}
\tablenotetext{2}{Continuum flux in units of $10^{-14}$ erg cm$^{-2}$ s$^{-1}$ \AA$^{-1}$.}
\tablenotetext{3}{Emission line flux in units of $10^{-12}$ erg cm$^{-2}$ s$^{-1}$.}
\end{deluxetable}


\section{Analysis}

\subsection{Cross Correlations}

A main purpose of this and other monitoring campaigns is to measure the
dimensions of the gas distribution in the BLR of Seyfert galaxies.
This is done by cross correlating the line and continuum light curves
and determining the time lag between them.  Another goal is to study
the variability properties of the sources.  One of the methods we have
applied to our data is the cross-correlation algorithm suggested by
Gaskell \& Peterson (1987). In this method, the cross-correlation
function (CCF) is calculated twice for the two observed light curves
$a(t_{i})$ and $b(t_{i})$: once by pairing the observed $a(t_{i})$ with
the interpolated value $b(t_{i}-\tau )$, and once by pairing the
observed $b(t_{i})$ with the interpolated value $a(t_{i}-\tau )$. The
final CCF is taken to be the average of these two.  No extrapolation
was used to avoid introducing artificial data, and the two last
observed points of each light curve were omitted because of the large
separation between them and the rest of the light curve. Linear and
spline interpolation gave similar results.

A major disadvantage of such interpolation methods is the lack of
rigorous error estimates for the CCF and the deduced lag. One way to
estimate the error is by assuming a certain BLR geometry and using
simulations to find the significance of the measured lag (Maoz \&
Netzer 1989). Another way, suggested by Gaskell \& Peterson (1987), is
an analytic estimate for the uncertainty on the CCF peak position. This
is only a rough estimate which relies on specific assumptions, such as
uniform sampling of the data. We use it for lack of a better method.

An alternative way that avoids interpolation, to find the CCF and the
time lag with error estimates, is to use the Discrete Correlation
Function (DCF; Edelson \& Krolik 1988). The error estimate of this
algorithm has been questioned by several authors (e.g., White \&
Peterson 1994; \ \ Paper IV). \ \ An improved \ \ algorithm 

\onecolumn

\begin{figure}
\vspace{5 cm}
\epsscale{1.00}
\plotone{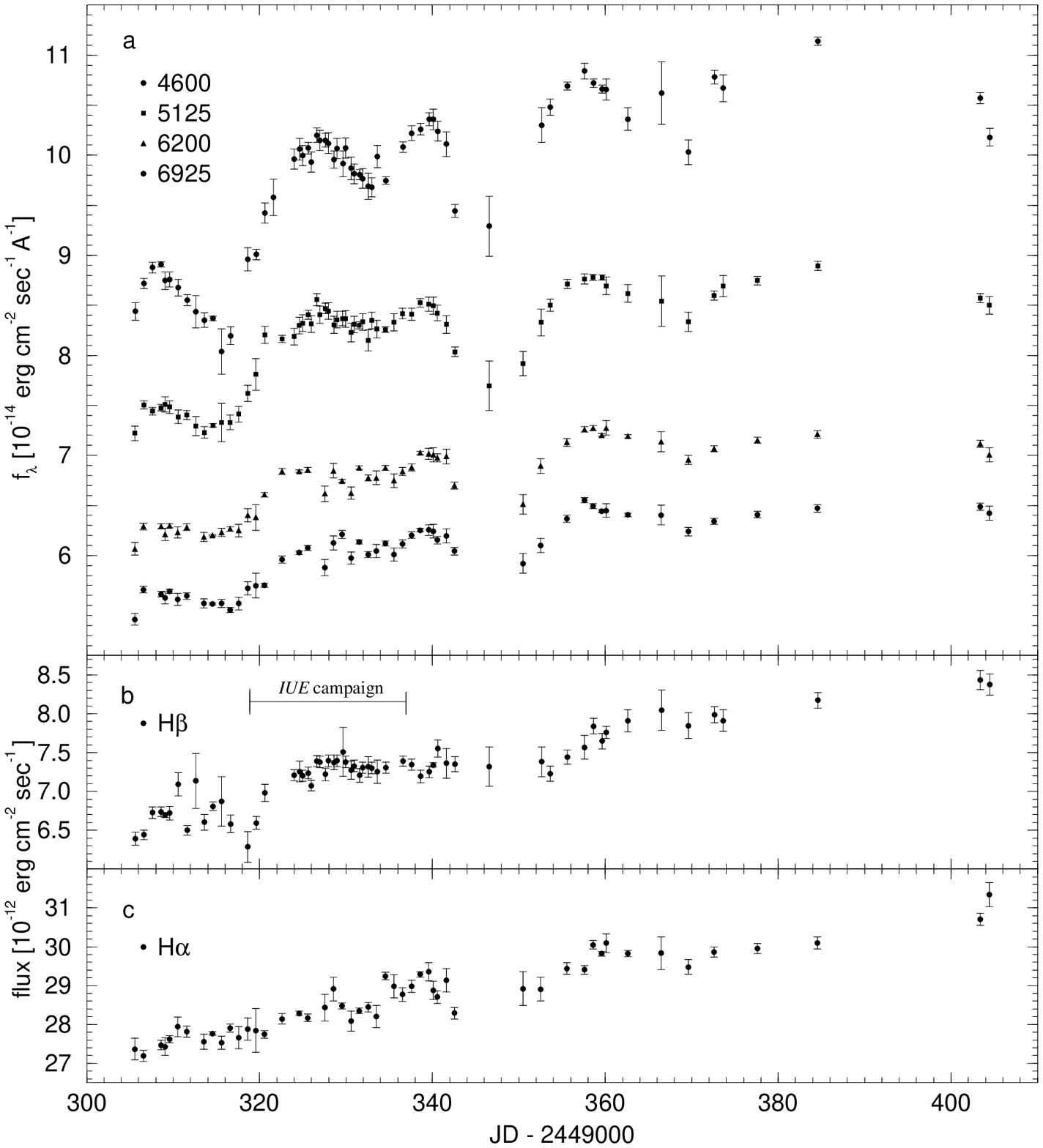}
\caption{(a) Four optical continuum light curves for NGC~4151. Wavelength
bands are given in upper left corner (in \AA). The ordinate
f$_{\lambda}$ is given in absolute units. No shifts are applied to the
top three light curves, and the 6925~\AA\ light curve is shifted by
-5$\times 10^{-15}$ erg cm$^{-2}$ s$^{-1}$ \AA$^{-1}$. (b) H$\beta$ light curve.
(c) H$\alpha$ light curve.}
\label{lc}
\end{figure}

\begin{figure}
\epsscale{0.90}
\plotone{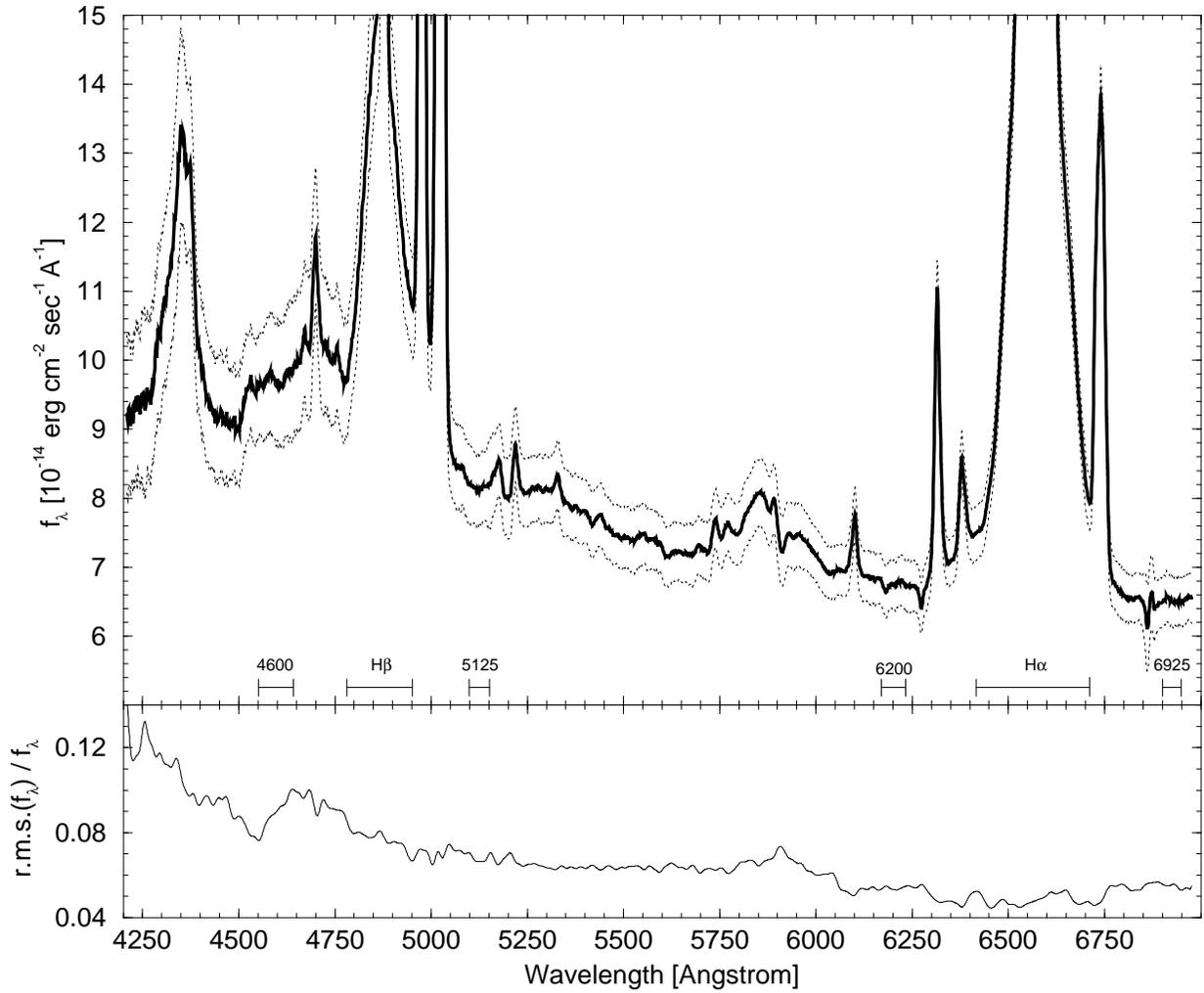}
\caption{Top panel: Mean spectrum of NGC~4151 from the Wise data set.
The dotted lines are the mean $+$ and $-$ the r.m.s. about the mean.
Residual telluric features are visible at 6280 and 6860 \AA .
The continuum and line measuring bins are marked.
Bottom panel: The smoothed ratio of the r.m.s. and the mean spectrum.
The variation of the light curves's amplitude and its wavelength
dependence can be seen from the larger variance at blue wavelengths.}
\label{spectrum}
\end{figure}

\twocolumn

\noindent was recently suggested
by Alexander (1996). This new approach applies Fisher's $z$
transformation to the correlation coefficient, and bins the DCF by
equal population bins rather than equal time bins. It results in a more
robust and statistically reliable method, the $Z$-transformed Discrete
Correlation Function (ZDCF). The ZDCF peak position and its errors are
estimated by a maximum likelihood method that takes into account the
uncertainty in the ZDCF points.

Discrete binning implicitly assumes that the spacing between the data
points is uncorrelated with their observing times. The NGC~4151 data
treated here pose a special problem in this respect. The galaxy's sky
position during this project allowed it to be observed only at the end
of the night. As a result, most of the JDs of the Wise set are at 0.6
of a day, whereas most JDs of the OSU set are at 0.0 day
(Table~\ref{con_lc}). Since the OSU data only extend over one eighth of
the monitoring period, time lags of $n+0.6$ days (where $n$ is an
integer), which cross correlate OSU with Wise points, strongly depend
on the behavior of the light curve during the OSU period, whereas time
lags of $n$ days, which cross correlate Wise with Wise points and OSU
with OSU points, reflect the overall behavior of the light curve. This
special sampling pattern resulted in spurious fluctuations between
consecutive ZDCF points. Such fluctuations disappeared only when the
ZDCF bin size was enlarged, or when the OSU data were omitted from the
light curves.  Since the first option significantly decreases the
number of ZDCF points, we use only the Wise data in the subsequent ZDCF
time series analysis. Below we present results from the ZDCF and the
interpolated CCF methods (where data points from all sets were used).
Both methods yield very similar results.

\begin{figure}[t]
\plotone{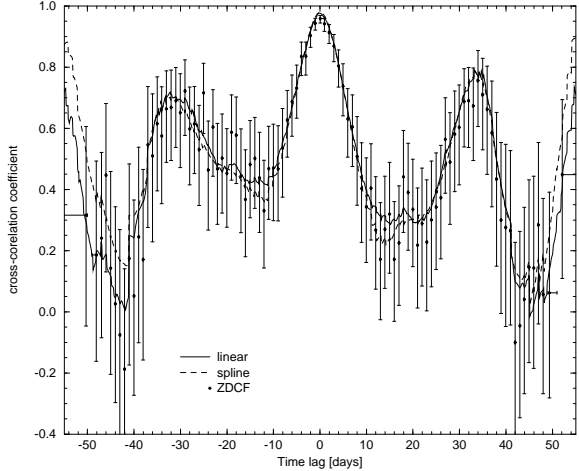}
\caption{CCFs of two continuum wavelength bands: 4600~\AA\ and
5125~\AA, with linear interpolation (solid line), spline interpolation
(dashed line), and ZDCF (error bars with filled circles). Note the good
agreement between the three methods.}
\label{con_ccf}
\end{figure}

Table~\ref{continuum_ccf} shows the properties of the various CCFs
calculated for the different continuum spectral bands.  In this table,
every wavelength band listed in a column is cross correlated with every
wavelength band listed in a row, such that if the wavelength band in
the row lags behind the wavelength in the column the time lag is
positive. For every two bands we calculated the CCF using the linear
interpolation (column ``lin'') and the ZDCF (column ``ZDCF'') methods.
For each CCF we give the time lag measured from the main peak, the
error on the time lag (as described above), the full width at half
maximum of the main peak (measured at half way between the peak's
maximum and minimum), and the correlation coefficient of the peak.  An
example of a CCF of two continuum wavelength bands is given in
Fig.~\ref{con_ccf}.  The diagonal in Table~\ref{continuum_ccf} gives
the auto-correlation function (ACF) results for all four continuum
wavelength bands. The ACFs are plotted in Fig.~\ref{con_acf}.

\begin{deluxetable}{lcccc}
\tablecolumns{5}
\tablecaption {Continuum CCFs \label {continuum_ccf}}
\tablewidth{0pt}
\tablehead{
\colhead{\phm{*****************}} & \colhead{4600~\AA}    & \colhead{5125~\AA}   & \colhead{6200~\AA}   & \colhead{6925~\AA} \nl
\colhead{} & \colhead{\ \ \ \ lin \ ZDCF} & \colhead{\ \ \ \ lin \ ZDCF} & \colhead{\ \ \ \ lin \ ZDCF} & \colhead{\ \ \ \ lin \ ZDCF}
}
\startdata
4600~\AA   &              &                &                      &            \nl
Time lag \dotfill	  & \phn 0.0 \phm{*} \phn {0.0}&  &      &         \nl
Time lag error \dotfill   &    \nodata \ \nodata   &            &         &         \nl
FWHM \dotfill             & 11.0 \phm{*} {10.0} &  &         &         \nl
Peak correlation \dotfill & 1.00 \phm{*} {1.00} &  &         &         \nl
\tablevspace{3mm}
5125~\AA           &       &            &         &         \nl
Time lag \dotfill	  & \phn 0.1 \phm{*} \phn {0.0} & \phn 0.0 \phm{*} \phn  {0.0}&  &      \nl
Time lag error \dotfill   & $^{+0.9}_{-0.9}$ \phm{*} \ $^{+0.9}_{-0.8}$ &   \nodata \ \nodata  &     &      \nl
FWHM \dotfill             & 11.5 \phm{*} {12.5} & 13.6 \phm{*} {14.4} &      &      \nl
Peak correlation \dotfill & 0.97 \phm{*} {0.96} & 1.00 \phm{*} {1.00} &     &      \nl
\tablevspace{3mm}
6200~\AA &           &            &         &         \nl
Time lag \dotfill	  & \phn 0.4 \phm{*} \phn {0.0} & \phn 0.4 \phm{*} \phn {1.0}& \phn 0.0 \phm{*} \phn {0.0}&      \nl
Time lag error \dotfill   & $^{+1.0}_{-1.0}$ \phm{*} \ $^{+0.9}_{-0.6}$ & $^{+0.9}_{-0.9}$ \phm{*} \ $^{+1.0}_{-1.6}$ &   \nodata \ \nodata   & \nl
FWHM  \dotfill            & 11.7 \phm{*} {10.5} & 11.9 \phm{*} {11.3} & 11.4 \phm{*} {12.3} &      \nl
Peak correlation \dotfill & 0.96 \phm{*} {0.95} & 0.97 \phm{*} {0.97} & 1.00 \phm{*} {1.00} &      \nl
\tablevspace{3mm}
6925~\AA &           &            &         &         \nl
Time lag \dotfill	  & \phn 0.6 \phm{*} \phn {0.0}& \phn 0.5 \phm{*} \phn {1.0}& \phn 0.1 \phm{*} \phn {0.0}& \phn 0.0 \phm{*} \phn {0.0} \nl
Time lag error \dotfill   & $^{+1.2}_{-1.2}$ \phm{*} \ $^{+1.4}_{-0.5}$ & $^{+1.1}_{-1.1}$  \phm{*} \ $^{+1.7}_{-1.2}$ & $^{+1.2}_{-1.2}$ \phm{*} \ $^{+0.9}_{-0.6}$ & \nodata \ \nodata \nl
FWHM   \dotfill           & 12.9 \phm{*} {12.6} & 12.4 \phm{*} {13.1} & 11.5 \phm{*} {11.5} & 11.0 \phm{*} {13.0} \nl
Peak correlation \dotfill & 0.97 \phm{*} {0.95} & 0.96 \phm{*} {0.96} & 0.98 \phm{*} {0.97} & 1.00 \phm{*} {1.00} \nl
\enddata
\tablenotetext{}{See explanation in text.}
\end{deluxetable}

\begin{deluxetable}{lccc}
\tablecolumns{4}
\tablecaption {Emission line CCFs \label {line ccf}}
\tablewidth{0pt}
\tablehead{
\colhead{\phm{*****************}} &  \colhead{5125~\AA}       &     \colhead{H$\alpha$}       &      \colhead{H$\beta$}    \\
\colhead{}       &  \colhead{\ \ \ lin \ {ZDCF}} & \colhead{\ \ \ lin \ {ZDCF}} & \colhead{\ \ \ lin \ {ZDCF}}
}
\startdata
H$\alpha$         &           &            &       \nl
Time lag \dotfill	  & \phn 0.6 \phm{*} \phn {1.0} & \phn 0.0 \phm{*} \phn {0.0}&           \nl
Time lag error  \dotfill  & $^{+1.7}_{-1.7}$  \phm{*} \ $^{+5.0}_{-1.9}$ &   \nodata \ \nodata     & \nl
FWHM       \dotfill       & 11.4 \phm{*} \phn {7.4} &   \nodata \ \nodata     &           \nl
Peak correlation \dotfill & 0.87 \phm{*} {0.89} & 1.00  \phm{*} {1.00} &           \nl
\tablevspace{3mm}
H$\beta$          &           &            &                  \nl
Time lag \dotfill	  & \phn 2.7 \phm{*} \phn {1.0} & \phn 0.6 \phm{*} \phn {0.1}& \phn 0.0 \phm{*} \phn {0.0}      \nl
Time lag error  \dotfill  & $^{+1.3}_{-1.3}$ \phm{*} \ $^{+2.7}_{-1.2}$ & $^{+5.2}_{-5.2}$  \phm{*} \ $^{+3.7}_{-11}$  & \nodata \ \nodata  \nl
FWHM       \dotfill       & 14.1 \phm{*}  {16.1} & 13.5 \phm{*} {22.5} & 10.8 \phm{*} {10.4}       \nl
Peak correlation \dotfill & 0.88 \phm{*}  {0.83} & 0.90 \phm{*} {0.83} & 1.00 \phm{*} {1.00}       \nl
\enddata
\tablenotetext{}{See explanation in text.}
\end{deluxetable}

Table~\ref{line ccf} lists  the properties of the CCFs of the 5125~\AA\
continuum light curve with the emission line (H$\alpha$ and H$\beta$)
light curves, as well as the CCF of the emission lines with each other.
This continuum band (which was observed in both the B-side and the
R-side of the Wise set) was chosen since it is the most reliable one.

\begin{figure}[t]
\vspace{2.6 cm}
\plotone{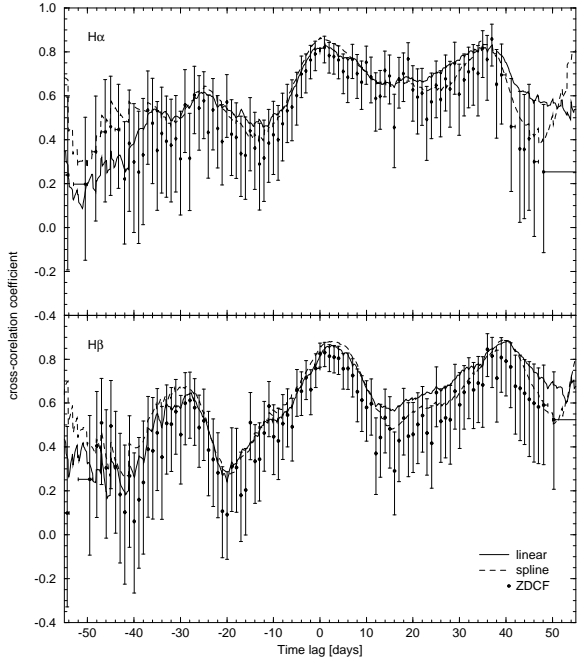}
\caption{CCFs of the 5125~\AA \ continuum band with H$\alpha$ (top
panel) and H$\beta$ (bottom panel). Notation is the same as in
Fig.~\protect\ref{con_ccf}. Note the time lag of $\sim 0-3$ days
indicated by the main peak.}
\label{Hb_ccf}
\end{figure}

The H$\alpha$ light curve (Fig.~\ref{lc}c) shows a small gradual rise
of $\sim12\%$ throughout the entire campaign. No significant signal
was found from cross correlating it with itself. In contrast, the
H$\beta$ light curve (Fig.~\ref{lc}b) shows, on top of the gradual
rise of $\sim30\%$, also some features that follow the continuum
variations. The CCFs of H$\alpha$ and H$\beta$ with the continuum light
curve are illustrated in Fig.~\ref{Hb_ccf}.

\subsection{Continuum Variability}

\begin{figure}
\plotone{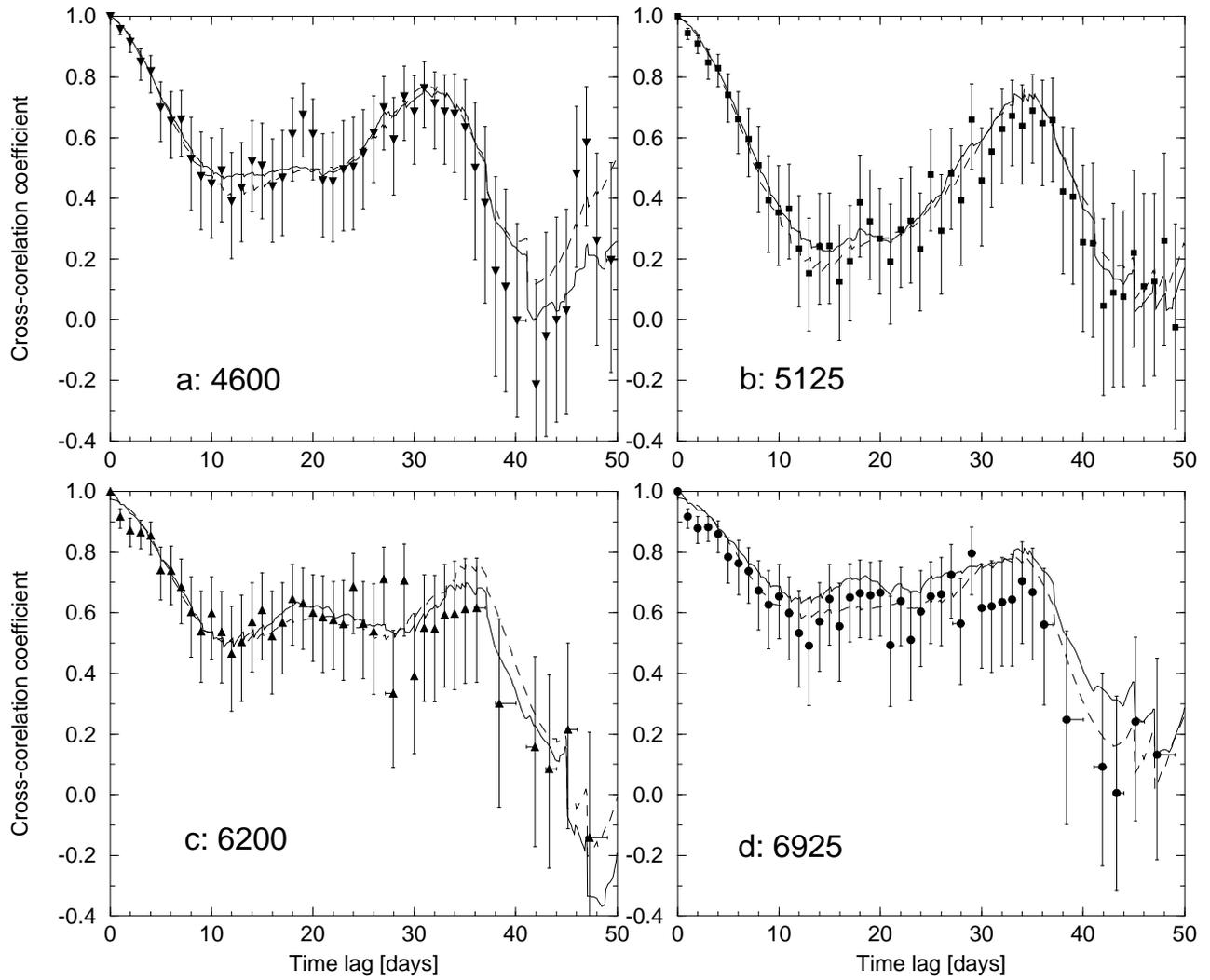}
\caption{ACFs of the four continuum bands presented in Table
\protect\ref{con_lc}. Notation is the same as in
Fig.~\protect\ref{con_ccf}. (a) 4600~\AA , (b) 5125~\AA , 
(c) 6200~\AA , and (d) 6925~\AA .}
\label{con_acf}
\end{figure}

\begin{figure}[t]
\vspace{1 cm}
\epsscale{1.40}
\plotone{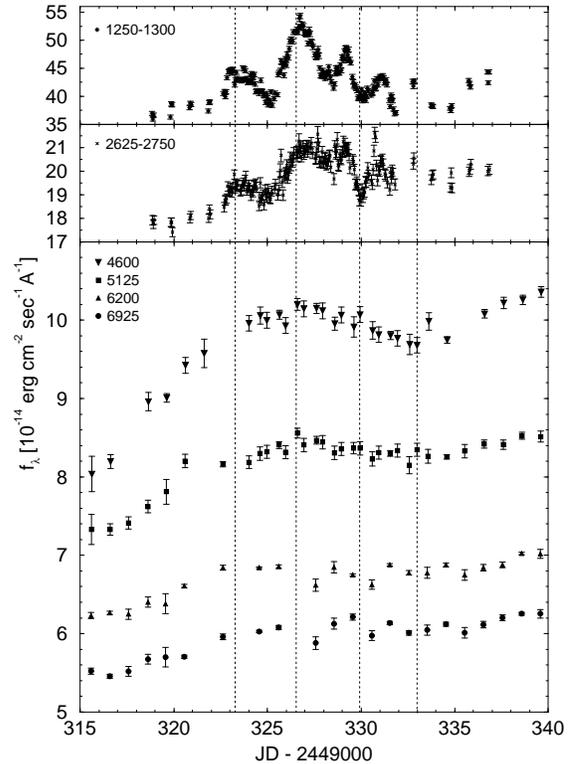}
\caption{Same as in Fig.~\protect\ref{lc}a for the period of the
$IUE$ campaign. Two UV continuum light curves are shown for
comparison with the optical continuum light curves.}
\label{UV_con}
\end{figure}

\begin{figure}
\epsscale{1.0}
\plotone{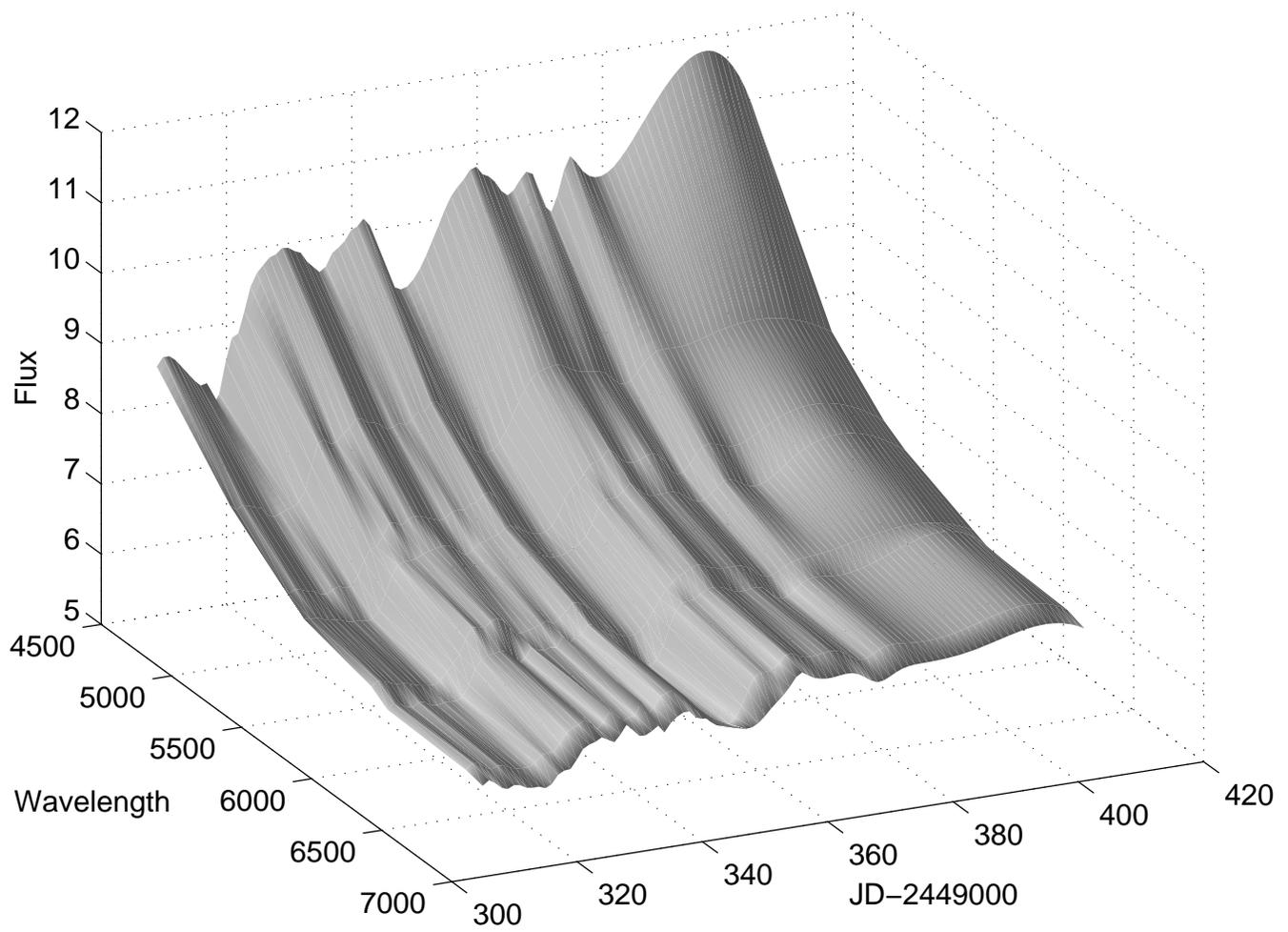}
\caption{Continuum variability of NGC~4151. Wavelength in units of \AA ,
flux in units of $10^{-14}$ erg cm$^{-2}$ s$^{-1}$ \AA$^{-1}$}
\label{3d_lc}
\end{figure}

Our monitoring of NGC~4151 for over two months with a temporal
resolution of 1 to 4 days resulted in light curves with errors of 1\%,
and can reveal variations as low as 3\%. The variability timescale,
defined as the FWHM of the main ACF peak (Table~\ref{continuum_ccf} and
Fig.~\ref{con_acf}), is $\sim13$ days. The measured continuum
variations are of 17\% to 35\%, with amplitude decreasing towards
longer wavelengths (Fig.~\ref{lc}a).  This trend extends to
shorter wavelengths, as seen by comparing the variability at UV
wavelengths (Paper I) with the optical light curves presented here
(Fig.~\ref{UV_con}).  Fig.~\ref{spectrum}, which shows the average and
r.m.s.  spectrum of NGC~4151 from the Wise data set, illustrates this
variable amplitude (see the larger variance at blue wavelengths in
the bottom panel). Fig.~\ref{3d_lc} presents this phenomenon in three
dimensions: wavelength, JD, and flux.  Here we have made use of some
more continuum wavelength bands, and the surface was interpolated over
missing data points and smoothed.

\begin{figure}[t]
\epsscale{0.9}
\plotone{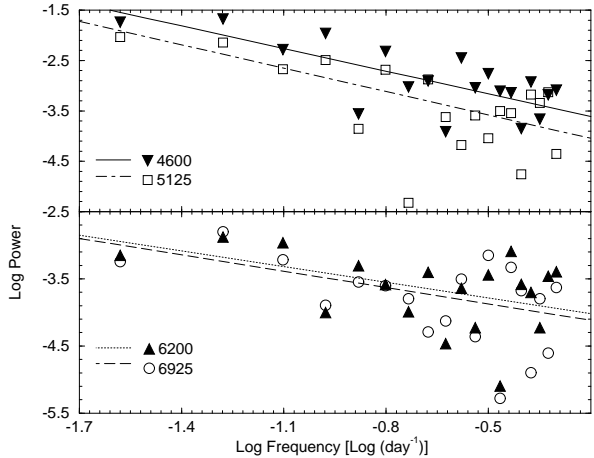}
\caption{Power Density Spectra for the four continuum wavelength bands
(calculated points and power-law fits). Note the decreasing
power in variability with increasing wavelength.}
\label{pspec}
\end{figure}

To quantify the variability amplitude, and its wavelength dependence,
we calculated the power spectra of the four continuum light curves.
Since the first 38 days of the monitoring period are evenly and almost
regularly sampled by the Wise data set, it is reasonable to apply a
discrete Fourier transform to these data. The power spectra are
presented in Fig.~\ref{pspec}. The decreasing power of the variability
with increasing wavelength is clear. A power-law fit, of the form
$PDS\propto f^{\alpha}$, to the first six points gives an index
$\alpha$ of $-1.5\pm0.9$ for \ \ the \ \ wavelength \ \ 
bands 4600~\AA\  \ \ and 5125~\AA , and 

\onecolumn

\twocolumn

\noindent $-0.8\pm0.6$ for the wavelength bands 6200~\AA\ and
6925~\AA . The small variation amplitude, especially at the longer
wavelengths, prevents us from determining a significant dependence of
variation timescale on wavelength.

The dependence of the variability amplitude on wavelength can be
explained by the varying relative contribution of starlight from the
underlying galaxy. We show this by using template spectra of Sab
galaxies from Coleman, Wu, \& Weedman (1980) and Kinney et al. (1996).
Based on previous observations of NGC~4151 (see Maoz et al.  1991) we
estimate the galaxy contribution to the spectrum through the Wise
aperture, in the continuum wavelength band of 4600~\AA, to be in the
range $(2-3.5)\times10^{-14}$ erg~cm$^{-2}$~s$^{-1}$~\AA$^{-1}$.
These numbers are confirmed by Peterson et al. (1995) who find the
starlight contribution through the Wise aperture to be
$2.2\times10^{-14}$ erg~cm$^{-2}$~s$^{-1}$~\AA$^{-1}$.  We normalize
the template spectra to these numbers and subtract the appropriate
values from the four continuum light curves. When using the maximum
value for the galaxy contribution at 4600~\AA\ we find that the
relative change in each light curve during the 14 days of intensive
monitoring is about 20\%, i.e., the differences in amplitude between
the light curves presented in Fig.~\ref{UV_con} disappear. The 20\%
variation amplitude is also similar to that of the UV continuum at
2625$-$2750~\AA\ (hereafter 2688~\AA ) but is different from that of
the 1250$-$1300~\AA\ (hereafter 1275~\AA ) light curve where the
relative variation is $\sim$35\%.

\begin{figure}[t]
\vspace{-0.5 cm}
\epsscale{1.40}
\plotone{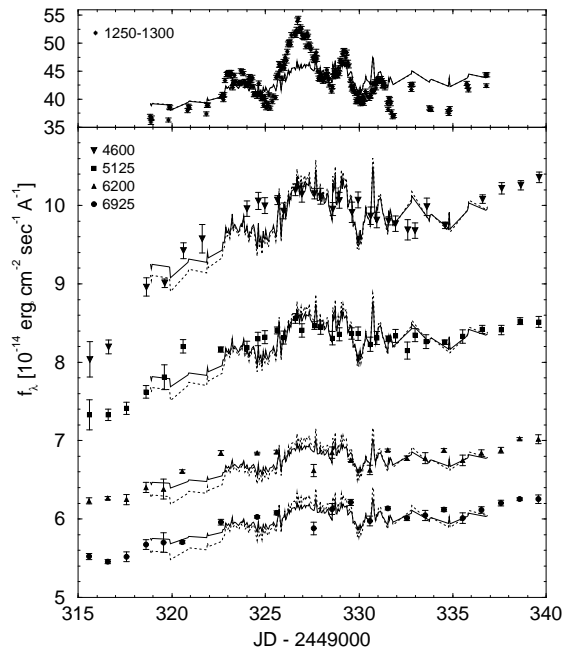}
\caption{Light curves from Fig.~\protect\ref{UV_con} superposed on the
linearly-interpolated 2688~\AA\ continuum light curve, after it was
scaled and diluted with a constant starlight flux according to the
galaxy contribution to the different wavelength bands: solid lines
correspond to a galaxy contribution at 4600 \AA\ of $3.5\times10^{-14}$
erg cm$^{-2}$ s$^{-1}$ \AA$^{-1}$, and dotted lines correspond to a
contribution of $2\times10^{-14}$ erg cm$^{-2}$ s$^{-1}$ \AA$^{-1}$.
Note the agreement between the scaled UV light curve and the optically
observed points (bottom panel), indicating that the decrease in
variability amplitude at larger wavelengths is mainly the result of
dilution by the constant galaxy light. The 1275~\AA\ UV light curve
(top panel) is qualitatively different from the scaled 2688 \AA\ light
curve.}
\label{fit_gal}
\end{figure}

We illustrate this result in Fig.~\ref{fit_gal}, where we present a
simulation of the continuum light curves based on the 2688~\AA\
continuum light curve. From the normalized galaxy template we separated
the total flux in this light curve into a galaxy contribution ($\sim
2\%$) and AGN contribution ($\sim 98\%$). The resulting AGN light curve was
scaled by a factor according to its part in the other
wavelength ranges and was added to the galaxy contribution at each
wavelength range. (The upper limit of the galaxy contribution at
4600~\AA , of $3.5\times10^{-14}$~erg~cm$^{-2}$~s$^{-1}$~\AA$^{-1}$,
gives the solid lines in Fig.~\ref{fit_gal}, and the lower limit, of
$2\times10^{-14}$~erg~cm$^{-2}$~s$^{-1}$~\AA$^{-1}$, gives the dotted
lines.) The observed data at all optical wavelengths show good
agreement with the galaxy-diluted UV light curve. The wavelength
dependence of the variability amplitude in the range
2700$-$7200~\AA\ can thus be explained by the different starlight
contribution at different wavelengths. The 1275~\AA\ UV continuum light
curve's amplitude cannot be explained using the starlight contribution
alone, as illustrated in the top panel of Fig.~\ref{fit_gal}.

From the CCFs of the different continuum wavelength bands with each
other, no significant lags are detected, i.e., the continuum varies in
phase at all optical wavelengths, to within 1 day.  There is no
apparent lag between the optical and UV light curves, as discussed
in~\ref{Emission_Line_Variability} and in Paper IV.

No clear conclusions result from comparison of the optical light curves
to the high-energy light curves (Paper III), since the high-energy
monitoring period was brief compared to the optical variation
timescale, and the number of data points is relatively small.

\onecolumn

\twocolumn

\subsection{Emission Line Variability}
\label{Emission_Line_Variability}

The H$\alpha$ light curve (Fig.~\ref{lc}c) shows a small gradual rise
throughout the entire campaign, and its ACF peak has a width of the
duration of the program. No variability timescale can be deduced from
this ACF. Cross correlation of H$\alpha$ with the 5125~\AA\ continuum
gives a time lag of $0-2$ days (Fig.~\ref{Hb_ccf} top panel). The
H$\beta$ ACF indicates a variability timescale similar to that of the
continuum (Table~\ref{line ccf}). Its CCF with the 5125~\AA\ continuum
(Fig.~\ref{Hb_ccf} bottom panel) suggests a lag, with respect to the
continuum light curve, of $0-3$ days. The CCF of the H$\alpha$ and the
H$\beta$ light curves shows no lag, implying that both emission lines
varied in phase.

We have performed cross correlations of the UV continuum light curves
(presented in Fig.~\ref{UV_con}) with the H$\alpha$ and H$\beta$
emission-line light curves. No significant time lags were found from
these CCFs' peaks. The strong variations in the UV continuum light
curves are of order 1$-$2 days and such rapid variations were not
detected in the optical emission-line light curves. This suggests a BLR
size larger than 1$-$2 lt-days, in which light-travel-time effects
smear out the rapid variations of the ionizing continuum. Since no
large timescale variations (of order 10 days) took place during the
UV monitoring period, no time lag between the UV light curves and the
optical emission lines could be found.

Detailed analysis and cross correlation of the optical light curves
with the UV light curves (Paper I), for the $IUE$ monitoring period,
are shown and discussed in Paper IV.  Here we discuss only the
correlation of the UV light curve with the entire optical period that
include 14 days of monitoring at the Wise Observatory, prior to the
beginning of the $IUE$ campaign.

\begin{figure}
\epsscale{0.9}
\plotone{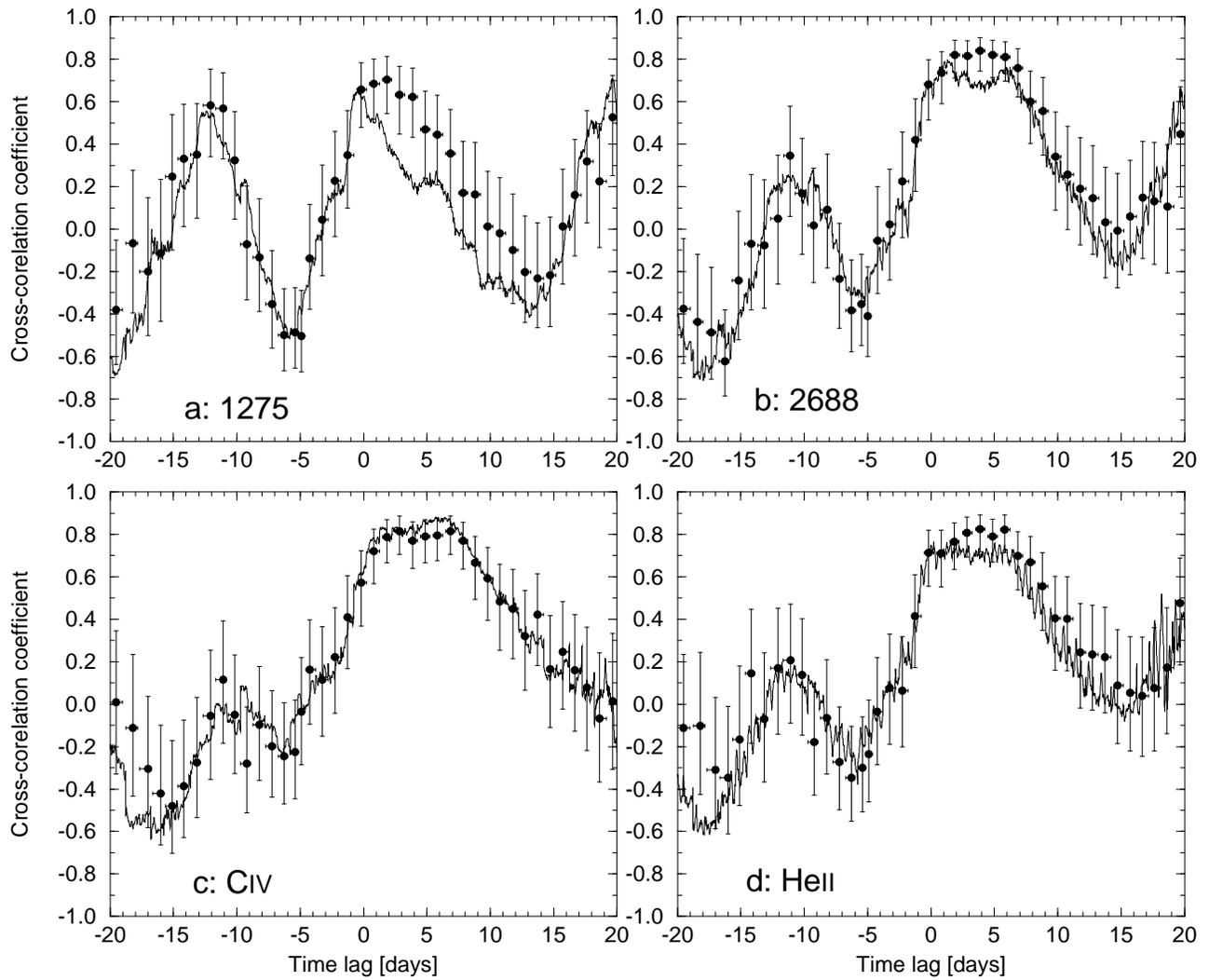}
\caption{CCFs of the entire optical continuum 5125~\AA\ band with the (a)
1275~\AA\ UV continuum, (b) 2688~\AA\ UV continuum, (c) \Civ\ $\lambda1549$
line, and (d) \Heii\ $\lambda1640$ line.  Notations are the same as in
Fig.~\protect\ref{con_ccf}. The spline interpolation CCF is not shown.}
\label{opt_UV_CCF}
\end{figure}

Cross correlation of the entire optical 5125~\AA\ continuum light curve
with the UV 1275~\AA\ and 2688~\AA\ continuum and the UV lines
\Civ\ $\lambda1549$ and \Heii\ $\lambda1640$ (Paper I) are shown in
Fig.~\ref{opt_UV_CCF}. All three CCFs, of 2688~\AA , \Civ , and \Heii
, show the same feature of a broad maximum, ranging from 0 to 6 days.
We believe that this apparent positive delay is an artifact due to
the finite duration of the monitoring campaign.
This broad feature is probably a result of the optical-continuum
behavior prior to the UV monitoring: these three UV light curves show a
gradual rise followed by period of constant flux and the
optical-continuum light curve shows a gradual rise prior to the
beginning of the UV monitoring. Those similar shapes result in high
correlation $(\sim 0.8)$ over a period of a few days.

We have carried out two tests of this hypothesis. We have extrapolated
the UV 2688~\AA\ continuum to the period prior to the $IUE$ monitoring
based on the shape of the optical continuum. This was done by finding
several 5125~\AA\ continuum data points which coincide in time with
several 2688~\AA\ continuum data points. The fluxes of those pairs of
points were fitted with a linear function which was then used to scale
the 5125~\AA\ light curve prior to the $IUE$ campaign to the
2688~\AA\ level. This extrapolated UV light curve was cross correlated
with the optical continuum and with the \Civ\ line light curves, and
gave a CCF peak at zero lag.  The second test is cross correlation of
the optical and UV light curves using only the part of the optical

\onecolumn

\twocolumn

\noindent light curve which overlaps the $IUE$ campaign. As shown and discussed
in Paper IV \S 3.4, the results are narrower peaks CCFs
consistent with a zero time lag. Thus, both tests suggest that the time
lags between the optical continuum and the UV line and continuum light
curves are consistent with zero. The last conclusion explains the
narrow peak (0-2 days) CCF of the 5125~\AA\ continuum with the
1275~\AA\ continuum shown in Fig.~\ref{opt_UV_CCF}a.  The zero time lag
is, indeed the lag and the difference between this CCF and others shown
in Fig~\ref{opt_UV_CCF} are due to the fact that the
1275~\AA\ continuum has more features in its light curve than the other
UV light curves.

\subsection{Comparison with Past Results}

Maoz et al. (1991) monitored NGC~4151 in 1988 for a period of 8 months
with an average sampling frequency of once every 4 days.  They found
continuum variations of $\sim$20\%, similar to the 1993 amplitude, with
a typical timescale of $\sim$28 days, about twice what is found in the
present campaign ($\sim$13 days).  They also found that both H$\alpha$
and H$\beta$ followed the continuum variations and were able to
determine time lags of $9\pm2$ days with respect to the optical
continuum, for the two emission-line light curves. In this work we find
that the H$\alpha$ and H$\beta$ variations lag behind the optical
continuum variations by $0-3$ days.  In both campaigns the H$\beta$
response to the continuum variations is larger than the H$\alpha$
response.

To \ investigate \ whether these \ discrepancies in \ emission-line lags are
the result of the different continuum variability timescales or of
different analysis methods, we have applied the auto-correlation
methods described above to the Maoz et al. (1991) data. We confirm that
when using the interpolation methods (linear and spline) the variation
timescale is $\sim$28 days. However, using the ZDCF we find a variation
timescale of $\sim$15 days.  This difference can be explained by the
fact that the 1988 monitoring program had several large gaps which,
when interpolated, increased the correlation between the continuum and
line light curves at larger times.  Using the data from the last 96
days of the 1988 campaign, which are more regularly sampled and have
only one large gap, we find that both the interpolation and the ZDCF
methods yield the same variability timescale of $\sim$15 days.
NGC~4151 therefore had similar optical continuum variability behavior
in 1988 and in 1993.

We have also performed cross correlation of the emission lines and the
continuum of the 1988 data. All cross-correlation methods yield the
same results for the time lag, $\sim 8$ days, consistent with the lag
found by Maoz et al. (1991).  The lag remains the same when only the
last 96 points of the 1988 data are used. We conclude that the
differences in H$\alpha$ and H$\beta$ lags behind the continuum between
1988 and 1993 are not the results of different methods of analysis.

Finally we have checked by way of simulations whether the small lag we
measure for the emission lines in 1993 could be the result of the
particular form of the continuum variation in 1993, which consisted
mainly of a monotonic rise.  Assuming a spherical BLR shell of inner
radius 2 lt-days and outer radius 30 lt-days (found by Maoz et al. to
best fit their results), we calculated the emission-line light curve
produced by such a geometry when driven by the spline-interpolated
4600~\AA \ continuum light curve from the present campaign. We then
sampled these two light curves at the same epochs of our actual
observations, and added to the sampled points a noise similar to the
errors in our light curves. We applied to these light curves the
various cross-correlation techniques discussed above, and found the
CCFs to have broad peaks at lags of $\sim 10$ days.  Thus the above
thick shell geometry, with an assumed ionizing continuum similar to the
4600~\AA\ light curve of 1993, gives a lag which is much larger than
that which we measure. The small time lags found in the 1993 data are
probably not an artifact of the particular continuum behavior we
observed.

If the above differences in emission-line lag are real, they may be
related to the different state of NGC~4151 which, at the time of the
1993 campaign, was in an ``active'' state, in contrast to its lower
flux level during the Maoz et al. monitoring campaign of 1988. (The
optical continuum and line fluxes in the present monitoring campaign
are about a factor of two higher than in 1988; see also Oknyanskij,
Lyutyi, \& Chuvaev 1994.)  Evidence for a changing lag has been shown
by Peterson et al. (1994) in the one other Seyfert galaxy that has been
intensively monitored for several years, NGC~5548. A possible physical
explanation for the change in the lag is a real change in the BLR gas
distribution between 1988 and 1993. Such a change is, in principle,
possible considering the scales and velocities present in the nucleus
of NGC~4151 which yield a dynamical time scale of $\sim 3$ years (see
also Wanders 1994). An alternative explanation is that during the
present campaign, the optical continuum did not properly represent the
behavior of the ionizing continuum. In particular, if the variability
timescale of the {\em{ionizing}} continuum driving the lines is
different between the two campaigns, the resulting time lags for the
same BLR geometry can be very different (Netzer \& Maoz 1990; Netzer
1990). Some evidence for this is seen in the differently-shaped light
curves of the 1275~\AA\ and the 2688~\AA\ continua
(Fig.~\ref{fit_gal}). This possibility is examined further in Paper
IV.

\section{Summary} 
		
We have presented optical-band results of an intensive two-month
spectrophotometric monitoring campaign of the Seyfert galaxy NGC~4151,
with a typical temporal resolution of one day. The main results of this
campaign are as follows.

1. The continuum variations are between 17\% and 35\%, with decreasing
amplitude towards longer wavelengths. The broad H$\alpha$ line flux
varied by $\sim12\%$ and the broad H$\beta$ flux by $\sim30\%$.

2. The decreasing continuum variability found at longer wavelengths can
be explained by the varying contribution of starlight from the
underlying galaxy. The exception to this is the far-UV
1275~\AA\ continuum where the variations must be intrinsically larger.

3. The various optical continuum bands vary in phase, with a lag of $<$
1 day. The typical continuum variability timescale is $\sim13$ days and
is similar at all optical wavelength bands. The variability amplitude
and timescale are similar to those observed in the past in this
object.

4. No evidence for a time lag between the optical continuum and the UV
continuum and emission lines was found. This may be partially the
result of the short duration of the $IUE$ campaign. Paper~IV gives
details on interband phase lags derived from the CCFs.

5. The H$\alpha$ and H$\beta$ light curves follow roughly the continuum
variations and lag them by $0-3$ days, in contrast to past results
where a time lag of 9$\pm$2 days was found. This may be related to a
different variability timescale of the {\em{ionizing}} continuum, or to
a real change in the BLR gas distribution in the 5.5 years interval
between the two campaigns.

\vspace{1cm}

\acknowledgments We would like to thank P.~Albrecht, M.~Dietrich,
J.~Huchra, Yu.F. Malkov, S.L.~Morris, V.I.~Pronik, S.G.~Sergeev,
J.C.~Shields, and B.J.~Wilkes for contributing their data to this
project. We are also grateful to John Dan, of the Wise Observatory
staff, for his dedicated assistance with the observations. I. Wanders
is thanked for providing his code for optimum [\OIII ] scaling.  The
work of the UC Berkeley team was supported by NSF grant AST-8957063 to
A.V.F.

\end{document}